\documentclass[a4paper,fleqn,useAMS,usenatbib]{mnras}

\pdfoutput=1

\usepackage{mathptmx}

\usepackage[T1]{fontenc}
\usepackage{ae,aecompl}

\usepackage[dvips]{graphicx}
\usepackage{amsmath}
\usepackage{amsfonts}
\usepackage{amssymb}
\usepackage{comment}
\usepackage[dvipsnames]{xcolor}
\usepackage[%
        ]{hyperref}
        
\hypersetup{
	colorlinks=true,	
	urlcolor=MidnightBlue,		
	pdfpagelabels=true,
	hypertexnames=true,
	plainpages=false,
	naturalnames=true,
	pdftitle={Observational biases for transiting planets},    
	pdfauthor={Kipping \& Sandford},     
	linkcolor=WildStrawberry,          
	citecolor=ForestGreen,        
}

\newcommand{\Kepler}{{\it{Kepler}}}

\newcommand{\Iout}{\overline{I_{\mathrm{out}}}}
\newcommand{\Iin}{\overline{I_{\mathrm{in}}}}

\newcommand{\Tobs}{T_{\mathrm{obs}}}

\newcommand{\SNR}{\mathrm{SNR}_{\mathrm{single}}}
\newcommand{\SNRm}{\mathrm{SNR}_{\mathrm{multi}}}

\newcommand{\tdet}{\hat{d}}
\newcommand{\tran}{\hat{b}}

\newcommand{\pdf}{\mathrm{Pr}}

\title[Observational biases for transiting planets]{Observational biases for transiting planets}
\author[Kipping \& Sandford]{David M. Kipping$^{1}$\thanks{E-mail:
\href{mailto:dkipping@astro.columbia.edu}{dkipping@astro.columbia.edu}} and Emily Sandford$^{1}$\\
$^{1}$Department of Astronomy, Columbia University, 550 W 120th Street, New York NY 10027, USA}

\date{Accepted . Received ; in original form }

\pubyear{2016}

\begin{document}
\label{firstpage}
\pagerange{\pageref{firstpage}--\pageref{lastpage}}
\maketitle

\begin{abstract}

Observational biases distort our view of nature, such that the patterns we see within a surveyed population of interest are often unrepresentative of the truth we seek. Transiting planets currently represent the most informative data set on the ensemble properties of exoplanets within 1 au of their star. However, the transit method is inherently biased due to both geometric and detection-driven effects. In this work, we derive the overall observational biases affecting the most basic transit parameters from first principles. By assuming a trapezoidal transit and using conditional probability, we infer the expected distribution of these terms both as a joint distribution and in a marginalized form. These general analytic results provide a baseline against which to compare trends predicted by mission-tailored injection/recovery simulations and offer a simple way to correct for observational bias. Our results explain why the observed population of transiting planets displays a non-uniform impact parameter distribution, with a bias towards near-equatorial geometries. We also find that the geometric bias towards observed planets transiting near periastron is attenuated by the longer durations which occur near apoastron. Finally, we predict that the observational bias with respect to ratio-of-radii is super-quadratic, scaling as $(R_P/R_{\star})^{5/2}$, driven by an enhanced geometric transit probability and modestly longer durations.

\end{abstract}

\begin{keywords}
methods: analytical --- methods: statistical --- eclipses --- planets and satellites: detection.
\end{keywords}

\section{Introduction}
\label{sec:intro}

The occurrence rate and properties of extrasolar planetary systems has been an
active area of astronomical research in recent years. These topics pertain
to our uniqueness in the cosmos and affect the technical requirements of future
instrumentation (e.g. see \citealt{dalcanton:2015}).

The primary \Kepler\ Mission, which was designed to be a statistical mission, has
provided a wealth of data to address these questions. By detecting the decrement 
in stellar brightness caused by planets passing in front of their stars, \Kepler\
has uncovered more than 4000 planetary candidates to date
\citep{burke:2015}. However, the transit method, whilst evidently highly successful, is
also plagued with several severe and well-known observational biases.
Specifically, two dominant forms of bias obscure our view of the true exoplanet population:
a geometric bias and a detection bias. For example, the former leads to a
bias towards detecting short-period planets \citep{beatty:2008} and the latter
leads to a bias towards larger planets.

Correcting for these biases is crucial when interpreting the \Kepler\ data to
make inferences about the underlying population. The most assured approach
for correcting for these biases is a full end-to-end numerical simulation of
the detection efficiency (e.g. see \citealt{petigura:2013}; \citealt{christiansen:2015};
\citealt{dressing:2015}). In such a simulation, one generates a
population of fake planets, creates corresponding fake time series with
transits, and pushes the fake time series through the detection pipeline of interest
to numerically compute the detection efficiency as a function of the input
parameters.

These numerical approaches are tailored to the specific mission of interest and
may be sensitive to the assumed population of injected planets. Nevertheless, 
this method is undoubtedly the most robust way to perform bias
corrections on complex data sets like \Kepler. Despite this, we argue that 
there is also great value in considering the problem of bias correction analytically. Analytic 
investigations can provide a deeper understanding of the underlying problem 
than possible through Monte Carlo simulations, which can often obscure the 
explanations of observed patterns. Moreover, analytic results can be
interpreted in any context, rather than being tailored to a specific mission.
This approach therefore complements the numerical work, offering insights into
which trends are instrument-specific and which are inherent to transit surveys.

Several previous works have investigated some of the biases affecting transit surveys analytically. For
example, \citet{barnes:2007} showed that the geometric bias of transits
favors the detection of eccentric transiting planets. \citet{burke:2008}
extended upon these ideas and showed how the transit duration is also affected
by orbital eccentricity, leading to an overall decrease in the detectability of eccentric planets.
\citet{beatty:2008} consider the overall detection yield from a transit survey
in an analytic framework, revealing the dependence on orbital period.

In these previous works, the transit light curves were assumed to be well-approximated
 by a simple box-like shape. Whilst a reasonable first
approximation, this assumption prohibits the investigation of V-shaped grazing
transits and is, as we show later, an assumption which could be relaxed without 
abandoning an analytic framework. Additionally, previous works tend to focus on
the expected detection yields from transit surveys (e.g. \citealt{beatty:2008}; 
\citealt{burke:2008}), as opposed to how the observed population parameters
are modified by the various biases. In \citet{eprior:2014}, we used conditional
probability to explore the distortion to the eccentricity distribution, but that
work was limited to just geometric bias (i.e., no detection bias) and also only
considered eccentricity and argument of periastron.

In this paper, we aim to build upon the aforementioned works to derive the
effect of both geometric and detection bias on the basic parameters observable
with transits. By using a trapezoidal transit model, we also investigate the
limit of grazing transits and use our model to explain previously observed
trends from the \Kepler\ transit survey.

The paper is organized as follows. In \S\ref{sec:SNR}, we derive an analytic 
expression for the SNR of a trapezoidal transit, including grazing geometries.
In \S\ref{sec:conditionals}, we show how making some approximations to this result 
allows us to write down a simple form for the joint distribution of the
transit parameters. We build upon this work in \S\ref{sec:advanced}, deriving
a more accurate set of expressions for the joint and marginalized 
distributions that result and extending the formalism to include occultations. 
Finally, we summarize our findings in \S\ref{sec:discussion}.

\section{Signal-to-Noise Ratio of a Transit}
\label{sec:SNR}

\subsection{SNR of a trapezoidal transit}

In order to analytically model detection bias, we begin by
deriving the SNR of the transit light curve as a function of the
basic transit parameters.

A transit is a decrease in the apparent brightness, or more specifically
the received intensity, $I(t)$, of a star as a function of time, $t$. The
received intensity may be defined as the number of photons received per
unit time. Away from transit events (`out-of-transit'), then, a
`vanilla' star displaying no changes in brightness would cause our
detector to receive an intensity

\begin{align}
I_{\mathrm{out}}(t) &= \Gamma,
\end{align}

where $\Gamma$ represents the nominal photon collection rate.
In order to actually measure the photon collection rate, we have to collect
photons over a specific time interval ($t_i \to t_f$) and then divide by
said interval. This is equivalent to calculating the mean of the function
$I(t)$ via

\begin{align}
\overline{I} =& \frac{ \int_{t=t_i}^{t_f} I(t) \mathrm{d}t }{ \int_{t=t_i}^{t_f} \mathrm{d}t }.
\label{eqn:Imean}
\end{align}

In the out-of-transit case, the above yields the simple result
$\overline{I_{\mathrm{out}}}=\Gamma$.

In the case of a star accompanied by a transiting planet, the intensity will be
periodically attenuated by the eclipse of the planet. Following 
\citet{carter:2008}, we approximate the shape of the transit light curve as a 
trapezoid in what follows. As was shown in fig.~1 of \citet{carter:2008},
four key times, $t_1<t_2<t_3<t_4$, as well
as a depth, $\delta$, define the shape of the transit:

\begin{equation}
I_{\mathrm{in}}(t)=\Gamma\left\{ \begin{array}{ll}
1-\delta & t_2 \le t \le t_3 \\
1-\frac{\delta (t-t_1)}{t_2-t_1} & t_1 < t < t_2 \\
1-\frac{\delta (t_4-t)}{t_4-t_3} & t_3 < t < t_4  \end{array}
       \right. .\label{eqn:Itransit}
\end{equation}

The fractional change in the measured intensity defines the `signal' in the
case of transits. We express the signal as

\begin{align}
S =& \frac{ \Iout - \Iin }{ \Iout },
\label{eqn:Sdef}
\end{align}

where the subscripts `in' and `out' denote in- and out-of-transit.
Evaluating $\Iin$ using Equations~(\ref{eqn:Imean}) and (\ref{eqn:Itransit}), then
substituting into Equation~(\ref{eqn:Sdef}) and replacing $(t_4-t_1)$
and $(t_3-t_2)$ with $T_{14}$ and $T_{23}$, respectively, yields

\begin{align}
S =& \Big( \frac{ T_{14} + T_{23} }{2 T_{14}} \Big) \delta.
\end{align}

In the limit of a box-like transit, we have $T_{14} \to T$ and $T_{23} \to T$,
which reproduces the classic result of $S = \delta$. 

Having calculated the `signal,' we now turn our attention to the `noise.' Since $\Iin$ and $\Iout$ are independent of one another, the error on $S$, which
we define as $\sigma_S$ (the `noise'), may be calculated via quadrature as

\begin{align}
\sigma_{S}^2 =& (\partial_{\Iout}S)^2\sigma_{\Iout}^2 + (\partial_{\Iin} S)^2 \sigma_{\Iin}^2,\nonumber\\
\qquad =& \frac{ \Iout^2 \sigma_{\Iin}^2 + \Iin^2 \sigma_{\Iout}^2 }{ \Iout^4 }.
\label{eqn:quadrature}
\end{align}

Since our detector is essentially a photon counter, the uncertainty on the terms
$\Iin$ and $\Iout$ may be calculated using Poisson statistics. Specifically, one
expects the uncertainty on the mean intensity $\overline{I(t)}$ to be given by

\begin{align}
\sigma_{\overline{I}} &= \frac{ \sqrt{ \int_{t=t_i}^{t_f} I(t) \mathrm{d}t } }{ \int_{t=t_i}^{t_f} \mathrm{d}t }.
\label{eqn:noisedef}
\end{align}

Evaluating $\sigma_{\Iin}$ and $\sigma_{\Iout}$ using Equation~(\ref{eqn:noisedef})
and then substituting the results into Equation~(\ref{eqn:quadrature}), we find

\begin{align}
\sigma_S &= \sqrt{ \frac{ (T_{14} - W \delta) (\Tobs + T_{14} - W \delta)}{ T_{14}^2 \Tobs \Gamma }},
\end{align}
	
where $W \equiv (T_{14}+T_{23})/2$ and $\Tobs$ is the duration over which the
out-of-transit intensity is observed. Finally, then, the SNR of a single trapezoidal
transit is

\begin{align}
\SNR &= \frac{ W \delta \sqrt{\Tobs \Gamma} }{ \sqrt{ (T_{14} - W \delta) (\Tobs + T_{14} - W \delta)} } .
\end{align}

In most instances, the out-of-transit baseline is much longer than the
in-transit observations in order to reduce the uncertainty in the transit
measurement. In this limit of $\Tobs \gg T_{14}$, we have

\begin{align}
\lim_{\Tobs \gg T_{14}}\SNR &= \frac{ W \delta \sqrt{\Gamma} }{ \sqrt{ T_{14} - W \delta } } 
\end{align}
	
In the small-planet case of $\delta \ll 1$ (appropriate even for the Jupiter-Sun system),
this becomes
	
\begin{align}
\lim_{\delta \ll 1}\lim_{\Tobs \gg T_{14}}\SNR &= \delta \sqrt{\Gamma} \frac{W}{\sqrt{T_{14}}}.
\label{eqn:SNR}
\end{align}

In what follows, $\Tobs \gg T_{14}$ and $\delta \ll 1$ are adopted as assumptions, and we do
not explicitly state them in subsequent expressions.

\subsection{Accounting for grazing transits}

Equation~(\ref{eqn:SNR}) provides a simple and practical estimate for the SNR of a
trapezoidal transit. However, it is framed in terms of the depth, $\delta$, which
is not equal to $p^2$ (the ratio-of-radii squared) in cases where
$(1-p)<b<(1+p)$, corresponding to so-called grazing transits. In such
instances, $\delta<p^2$ and requires calculating the area of partial overlap between
the planet's disc and that of the star. This is commonly done using the $\lambda$
function, defined in Equation~1 of \citet{mandel:2002} as

\begin{align}
&\lambda(p,\mathcal{S}) = \frac{1}{\pi} \Bigg( p^2 \kappa_0 + \kappa_1 - \sqrt{\frac{4\mathcal{S}^2-(1-\mathcal{S}^2-p^2)^2}{4}} \Bigg),
\label{eqn:lambda}
\end{align}

where

\begin{align}
&\kappa_0 \equiv \cos^{-1}\Big(\frac{p^2+\mathcal{S}^2-1}{2p\mathcal{S}}\Big),\\
&\kappa_1 \equiv \cos^{-1}\Big(\frac{1-p^2+\mathcal{S}^2}{2\mathcal{S}}\Big),
\end{align}

and $\mathcal{S}$ is the sky-projected planet--star separation in stellar radii.
Accordingly, we now express

\begin{equation}
\delta(p,b)=\left\{ \begin{array}{ll}
p^2               & 0 \le b < 1-p \\
\lambda(p,b) & 1-p \le b <1+p \\
    0                                          & 1+p \le b < \infty  \end{array}
       \right. .\label{eqn:delta}
\end{equation}
	
We may now modify the $\SNR$ expression (Equation~\ref{eqn:SNR}) to account for
grazing events as

\begin{equation}
\SNR=\left\{ \begin{array}{ll}
\Big( \frac{T_{14}+T_{23}}{2\sqrt{T_{14}}} \Big) p^2 \sqrt{\Gamma} & 0 \le b < 1-p \\
\Big( \frac{\sqrt{T_{14}}}{2} \Big) \lambda(p,b) \sqrt{\Gamma} & 1-p \le b < 1+p \\
    0                                          & 1+p \le b < \infty  \end{array}
       \right. ,\label{eqn:SNR2}
\end{equation}

where the grazing case is evaluated by appreciating that $T_{23}=0$ in such
instances. In the limit of a box-like transit, these expressions are equivalent
to Equation~19 of \citet{carter:2008}.

\subsection{Approximating the SNR equation}

In order to understand the relationship between SNR and impact parameter, $b$, we need
to rewrite Equation~(\ref{eqn:SNR2}) by evaluating the various durations. To this end, we
employ the ``one term'' transit duration function from \citet{investigations:2010} (Equation~15
 of that work) in what follows. We initially seek a simple and approximate formula, with the 
 full SNR expression of Equation~(\ref{eqn:SNR2}) serving as a useful comparison for
  the accuracy of proposed approximations.

We start by assuming that $p\ll1$, the small-planet approximation, for which the first
implication is that grazing transits contribute negligibly to the SNR function, and thus
we need only consider trapezoidal transits. Under this approximation, it also
true that one expects $W=(T_{14}+T_{23})/2\simeq\tilde{T}$, as defined in \citet{investigations:2010},
which reduces two of the duration terms to just one. We then proceed to make a small-angle
approximation on the inverse sine functions ($\sin^{-1}(x)\simeq x$) within the duration
terms, equivalent to assuming $a_R\gg1$, which is valid for the vast majority of
exoplanets. In total, then, we approximate $a_R\gg1$, $W \to \tilde{T}$ and $p\ll1$,
such that

\begin{align}
\lim_{a_R\gg1} \lim_{W \to \tilde{T}} \lim_{p\ll1} \SNR = \nu_{\mathrm{single}} (1-b^2)^{1/4} p^2,
\label{eqn:SNRapprox}
\end{align}

where we have defined

\begin{align}
\nu_{\mathrm{single}} &\equiv \Bigg(\frac{3}{G \pi^2}\Bigg)^{1/6} \rho_{\star}^{-1/6} P^{1/6} \Bigg(\frac{ (1-e^2)^{1/4} }{ (1+e\sin\omega)^{1/2}} \Bigg) \Gamma^{1/2}.
\end{align}

The accuracy of Equation~(\ref{eqn:SNRapprox}) versus the more accurate
Equation~(\ref{eqn:SNR2}) is illustrated later in Figs~\ref{fig:SNR_approx1}
and \ref{fig:SNR_approx2}.

Whilst the above is derived for single transits, one often deals with multiple events, in which case the SNR is expected
to increase with the square root of the number of events. Within a given
baseline of observations, $\Tobs$, we therefore expect the SNR to increase
by $\sqrt{\Tobs/P}$, since shorter-period planets fit more transits in
within $\Tobs$, such that

\begin{align}
\lim_{a_R\gg1} \lim_{W \to \tilde{T}} \lim_{p\ll1} \SNRm = \nu_{\mathrm{multi}} (1-b^2)^{1/4} p^2,
\label{eqn:SNRmulti}
\end{align}

where

\begin{align}
\nu_{\mathrm{multi}} &\equiv \Bigg(\frac{3}{G \pi^2}\Bigg)^{1/6} \rho_{\star}^{-1/6} P^{-1/3} \Bigg(\frac{ (1-e^2)^{1/4} }{ (1+e\sin\omega)^{1/2}} \Bigg) \sqrt{\Gamma \Tobs}.
\end{align}

\section{Conditional Probability Distributions}
\label{sec:conditionals}

\subsection{Probability of detection}
\label{sub:ramp}

In what follows, we use the symbol $\tdet$ to denote `transit detected.'
Let us group the various parameters in Equation~(\ref{eqn:SNRmulti}) into a
vector $\boldsymbol{\Theta} = \{p,b,P,\rho_{\star},e,\omega\}$. We define the
probability of detecting a transit conditioned upon these transit parameters
\emph{and} the fact that the system geometry permits the planet to transit at all as 
$\pdf(\tdet|\boldsymbol{\Theta},\tran)$, where $\tran$ is short-hand
for $b<1$. In what follows, a basic assumption of this work is that

\begin{align}
\pdf(\tdet|\boldsymbol{\Theta},\tran) \propto \SNRm,
\label{eqn:SNRproportional}
\end{align}

which states that the probability of detecting a transit is proportional to
its SNR. Whilst this statement is intuitive, it is worth pausing to consider its validity in the case of the most comprehensive transit
survey to date, \Kepler.

\begin{figure}
\begin{center}
\includegraphics[width=8.4 cm]{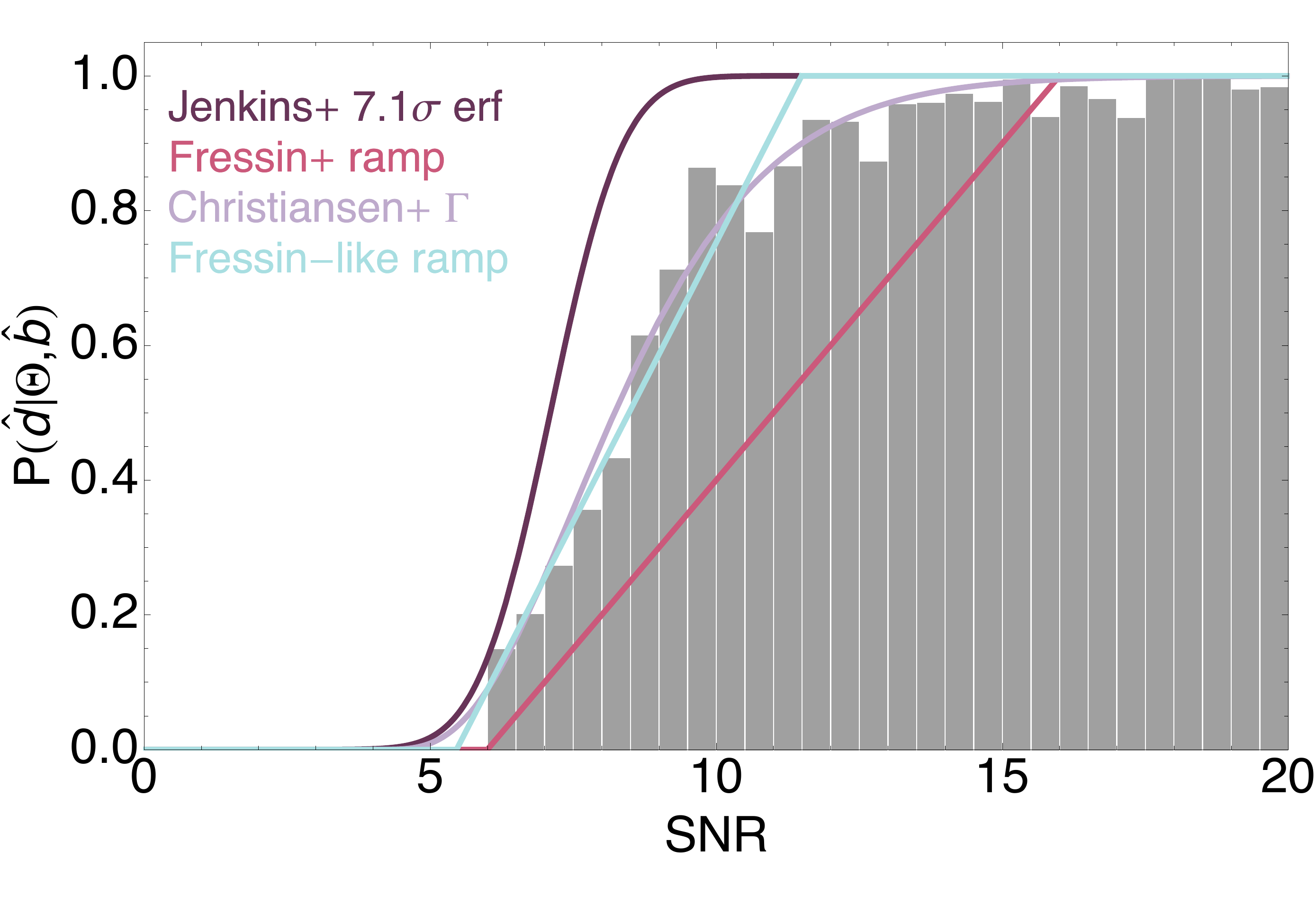}
\caption{
Probability of detecting an injected transit of a known SNR into the \Kepler\
pipeline for FGK dwarfs (data taken from \citealt{christiansen:2015}). The 
top-left key lists various models in chronological publication order, where the
last is derived here using updated parameters for the \citet{fressin:2013} ramp, 
which provides a reasonable description with a sufficiently analytically simple model
to make progress within this work.
}
\label{fig:christiansen}
\end{center}
\end{figure}

The formal criterion for a \Kepler\ detection is that the transit events display
an SNR exceeding 7.1 \citep{jenkins:2010}. This assumption implies that
the probability of detection, $\pdf(\tdet|\boldsymbol{\Theta},\tran)$, follows a 
cumulative normal distribution centred on 7.1 with a standard deviation of unity
\citep{christiansen:2015}, as shown in Fig.~\ref{fig:christiansen}.

In practice, this simple model has been found to be an unrealistic representation
of the \Kepler\ detections, with \citet{fressin:2013} arguing that modelling
$\pdf(\tdet|\boldsymbol{\Theta},\tran)$ as linearly proportional to SNR (the same assumption we make here) describes the data far better. However,
recently, \citet{christiansen:2015} used injection/recovery simulations to show
that $\pdf(\tdet|\boldsymbol{\Theta},\tran)$ is best described by a $\Gamma$ cumulative
distribution, as shown in Fig.~\ref{fig:christiansen}. Nevertheless, fitting
updated parameters to the \citet{fressin:2013} ramp model\footnote{For those
interested, we determined a ramp from SNR of 5.5--11.5 for the FGK dwarfs.}
conditioned on the FGK dwarf injection tests of \citet{christiansen:2015}
reveals that a linear model still provides an excellent description, as evident
in Fig.~\ref{fig:christiansen}. Due to its analytic convenience, reasonable accuracy, and general applicability
to other surveys, we therefore adopt the linearly proportional assumption throughout.

We also note that detection functions tend to truncate to unity probability
at a high SNR, and our model does not directly account for this truncation.
However, in such a regime, the transit surveys are complete, detection
bias plays no role, and one may use the geometric bias terms derived in this work
alone.

\subsection{Joint probability distribution of the transit parameters}

Using Bayes' theorem, the joint probability distribution of $\boldsymbol{\Theta}$,
conditioned upon $\tdet$ and $\tran$, is given by

\begin{align}
\pdf(\boldsymbol{\Theta}|\tdet,\tran) &\propto \pdf(\tdet|\boldsymbol{\Theta},\tran) \pdf(\boldsymbol{\Theta}|\tran),
\label{eqn:Theta_prior_scaling}
\end{align}

We take a brief aside to explore the $\pdf(\boldsymbol{\Theta}|\tran)$ term. From
\citet{eprior:2014}, Equation~2, the probability of a transit, defined as being
$b<1$, modified to replace $a_R \to \rho_{\star}$, is given by

\begin{align}
\pdf(\tran|\boldsymbol{\Theta}) &= \Bigg( \frac{3\pi}{ G } \Bigg)^{1/3} P^{-2/3} \rho_{\star}^{-1/3} \Bigg(\frac{1+e\sin\omega}{1-e^2}\Bigg).
\label{eqn:geometric}
\end{align}

The above essentially assumes $p\ll1$, since grazing events do not contribute as
`transits.' Using Bayes' theorem, then, one may write that

\begin{align}
\pdf(\boldsymbol{\Theta}|\tran) &\propto \pdf(\tran|\boldsymbol{\Theta}) \pdf(\boldsymbol{\Theta}),
\end{align}

and thus

\begin{align}
\pdf(\boldsymbol{\Theta}|\tdet,\tran) &\propto \pdf(\tdet|\boldsymbol{\Theta},\tran) \pdf(\tran|\boldsymbol{\Theta}) \pdf(\boldsymbol{\Theta}).
\label{eqn:Theta_prior_fullscaling}
\end{align}

where it is evident from Equation~(\ref{eqn:geometric}) that 
$\pdf(\tran|\boldsymbol{\Theta})$ is in fact independent of
$p$ and $b$. Since $\pdf(\tdet,\tran|\boldsymbol{\Theta}) 
\propto \pdf(\tdet|\boldsymbol{\Theta},\tran) \pdf(\tran|\boldsymbol{\Theta})$,
we may re-write Equation~(\ref{eqn:Theta_prior_fullscaling}) as

\begin{align}
\underbrace{\pdf(\boldsymbol{\Theta}|\tdet,\tran)}_{\text{overall dist.}} &\propto 
\underbrace{ \pdf(\tdet,\tran|\boldsymbol{\Theta})}_{\text{observational bias}}
\underbrace{\pdf(\boldsymbol{\Theta})}_{\text{intrinsic dist.}}.
\label{eqn:Theta_prior}
\end{align}

where $\pdf(\tdet,\tran|\boldsymbol{\Theta})$ is proportional to the product of
Equations~(\ref{eqn:SNRmulti}) and (\ref{eqn:geometric}), yielding

\begin{align}
\pdf(\tdet,\tran|\boldsymbol{\Theta}) &\propto \kappa (1-b^2)^{1/4} p^2,
\label{eqn:bias}
\end{align}

where we define

\begin{align}
\kappa &\equiv P^{-1} \rho_{\star}^{-1/2} (1+e\sin\omega)^{1/2} (1-e^2)^{-3/4}.
\label{eqn:kappa}
\end{align}

It is worth noting that in the above, all of the terms except $e$ and $\omega$ are
separable, and Equation~(\ref{eqn:bias}) can be expressed as

\begin{align}
\pdf(\tdet,\tran|\boldsymbol{\Theta}) &\propto 
\pdf(\tdet,\tran|P) 
\pdf(\tdet,\tran|\rho_{\star}) 
\pdf(\tdet,\tran|e,\omega)
\pdf(\tdet,\tran|b)
\pdf(\tdet,\tran|p).
\end{align}

If $\pdf(\boldsymbol{\Theta})$, the intrinsic distribution, is also separable, then the
result above implies that $\pdf(\boldsymbol{\Theta}|\tdet,\tran)$ is separable too.
In practice, the intrinsic distribution is unlikely to be truly separable (even if assumed
so for convenience). In particular, the distribution of ratio-of-radii, $p$, is likely
covariant with terms such as $\rho_{\star}$. This means that inverse sampling from the
overall prior will be hierarchically conditional, assuming it is even possible to
write down a closed form for such a distribution. For this reason, we argue that
inverse transform sampling is not a practical way of sampling the distribution.

\subsection{Using the conditional probability distributions as a prior}
\label{sub:droprhostar}

As a brief aside, we discuss here why generally we do not recommend blindly
using the derived conditional probabilities as priors in the analysis of
individual transiting planets. To centre the discussion, we first consider the
case of the mean stellar density, $\rho_{\star}$.

The likelihood function of a real transit fit exhibits a strong negative covariance
between $\rho_{\star}$ and $b$ \citep{carter:2008}. Our derived bias terms favour
a low $\rho_{\star}$ and a low $b$, i.e., they are positively covariant. Furthermore,
the exponential $\rho_{\star}$ bias quickly outweighs the softer $b$ bias, and thus,
when convolved with the likelihood effect, transit fits would tend to be driven towards
low $\rho_{\star}$ parameter space but high, near-grazing $b$. This means the $b$
bias is essentially washed out, and the overall fit is driven towards low
$\rho_{\star}$.

One reason why this is problematic is that, thanks to our knowledge of stellar
evolution, we understand that low-density stars will have larger
stellar radii, and therefore, the same sized planet is less likely to be detected
around a low-density than a high-density star. This problem could be resolved
if we actually knew the true joint (and covariant) intrinsic prior of $\rho_{\star}$
and $p$, which would correct for this effect. Of course, the intrinsic prior is generally
unavailable.

Since many of the transit parameters display covariance in the likelihood
function to some extent \citep{carter:2008}, particularly in the case of
realistic transits with limb darkening and sparse sampling, the above effect is
general. For these reasons, we do not advocate blindly using these distributions as 
priors in individual transit light-curve fits.

\section{Extending to Grazing Transits}
\label{sec:advanced}

\subsection{A more accurate SNR}

Based on the analytical form for the SNR of a trapezoidal transit and the
geometric bias of the transit method, we have derived the conditional
probability of the basic transit parameters for a planet detected to
transit, $\pdf(\boldsymbol{\Theta}|\tdet,\tran)$.

In order to do this, we made various approximations to the SNR expression
shown in Equation~(\ref{eqn:SNR2}). The result of these approximations
was a simple, multiplicative form of the joint distribution 
(Equation~\ref{eqn:bias}). Whilst this is undoubtedly useful, especially
for illustrating overall trends, we describe here a more accurate estimate.

The biggest problem with Equation~(\ref{eqn:bias}) is that $b=1$
has zero probability, but the SNR should not go to zero until $b=(1+p)$.
Therefore, our proposed simple form truncates a part of parameter space.
To derive a superior approximation, we return to Equation~(\ref{eqn:SNR2}).

\begin{figure}
\begin{center}
\includegraphics[width=8.4 cm]{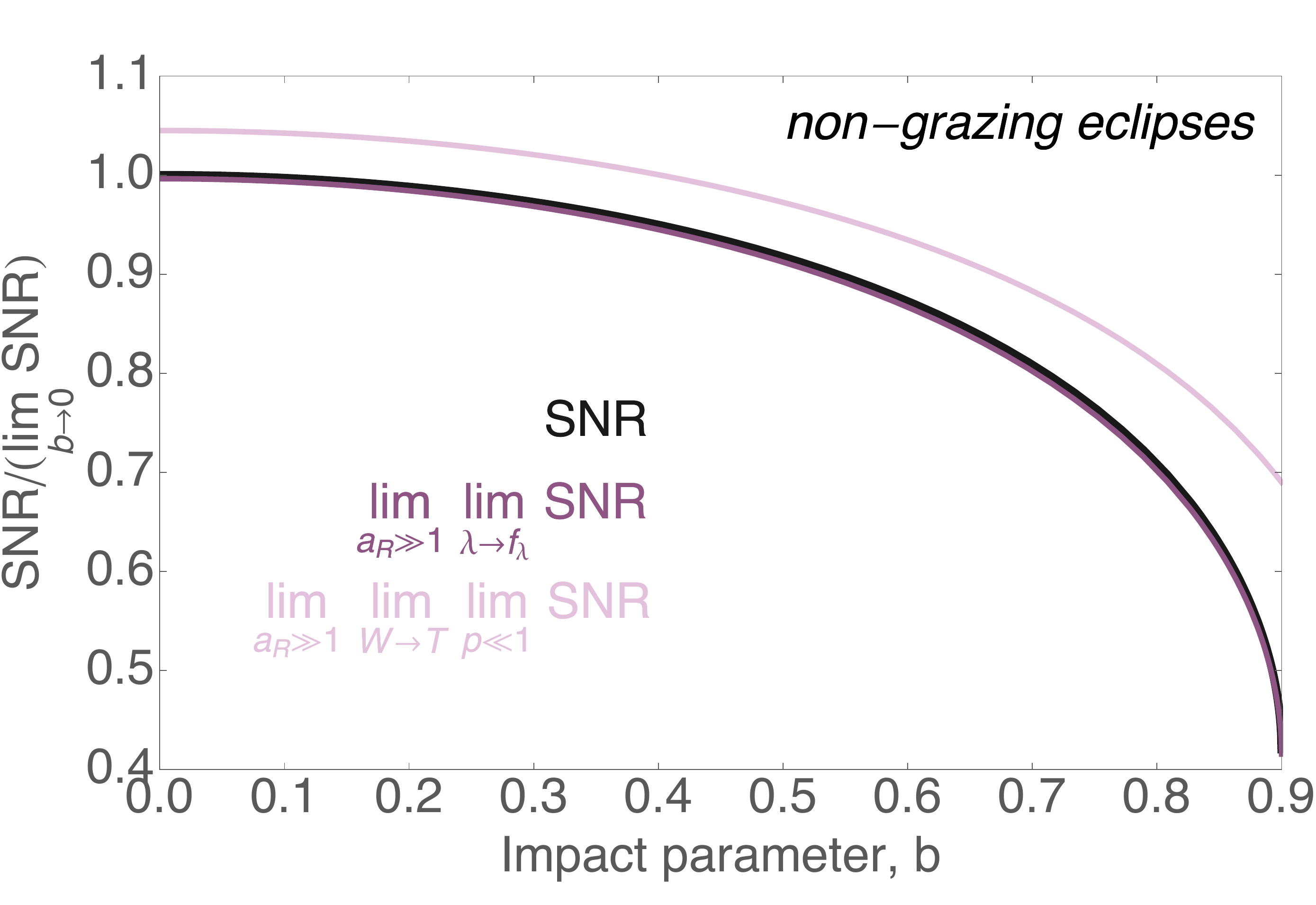}
\caption{
Comparison of three approximations (coloured lines) to the
full SNR expression (black) for non-grazing transits. We here adopt
$p=0.1$, $P=3$\,d, $a_R=10$, $e=0.1$ and $\omega=0$. Whilst
the most approximate form (Equation~\ref{eqn:SNRmulti}) only broadly reproduces the shape, assuming only $a_R\gg1$ (Equation~\ref{eqn:SNRin}) provides an excellent approximation.}
\label{fig:SNR_approx1}
\end{center}
\end{figure}

In the case of non-grazing events, we first tried approximating
$(T_{14} + T_{23})/2 \simeq \tilde{T}$ (see \citealt{investigations:2010}
for definitions). However, this approximation becomes increasingly inaccurate as
$b\to(1-p)$ (see Fig.~\ref{fig:SNR_approx1}). For a more accurate expression,
we only apply $a_R\gg 1$, allowing us to make small-angle approximations, to show 
that:

\begin{align}
\lim_{a_R\gg1} \SNRm^{\text{non-grazing}} =& \frac{\nu_{\mathrm{multi}}}{2} ((1+p)^2 - b^2)^{-1/4} p^2 \nonumber\\
\qquad & \Big( \sqrt{(1+p)^2 - b^2} + \sqrt{(1-p)^2 - b^2} \Big),
\label{eqn:SNRin}
\end{align}

where the superscript `non-grazing' defines the condition under
which the above is valid. In the limit of $p\to0$, 
this SNR expression reduces to the simpler approximation of 
Equation~(\ref{eqn:SNRmulti}).

In the case of a grazing event, we formally need to use the $\lambda$ 
function defined earlier in Equation~(\ref{eqn:lambda}):

\begin{align}
\lim_{a_R\gg1} \SNRm^{\text{grazing}} = \frac{\nu_{\mathrm{multi}}}{2} ((1+p)^2-b^2)^{1/4} \lambda.
\label{eqn:SNRgraz}
\end{align}

The $\lambda$ function (Equation~\ref{eqn:lambda}) can be approximated to
high accuracy using the $f_{\lambda}$ derived in Appendix~\ref{app:lambda}
(see Equation~\ref{eqn:flambda}). Using this approximation, we may write

\begin{align}
\lim_{a_R\gg1} \lim_{\lambda\to f_{\lambda}} \SNRm^{\text{grazing}} =& \frac{\nu_{\mathrm{multi}}}{2} ((1+p)^2-b^2)^{1/4} \nonumber \\
\qquad& \frac{(1+p-b)^2}{4} \Bigg( 1 - 2\log_4\Big[\frac{1+p-b}{2p}\Big] \Bigg).
\label{eqn:SNRgrazing}
\end{align}

\begin{figure}
\begin{center}
\includegraphics[width=8.4 cm]{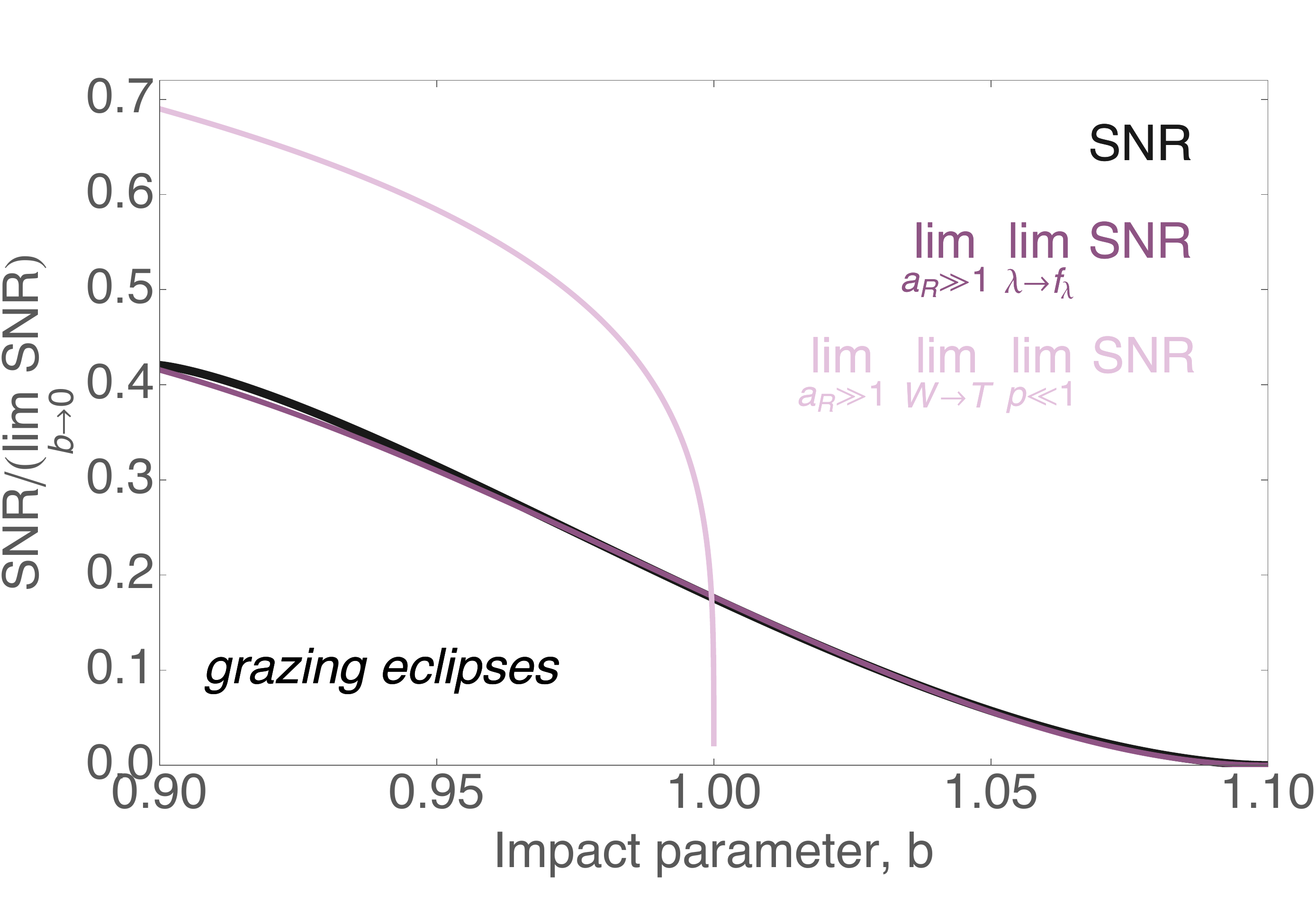}
\caption{
Comparison of three approximations (coloured lines) to the
full SNR expression (black) for grazing transits. We here adopt
$p=0.1$, $P=3$\,d, $a_R=10$, $e=0.1$ and $\omega=0$. The
most approximate form (Equation~\ref{eqn:SNRmulti}) poorly reproduces the shape, but assuming $a_R\gg1$ and $\lambda 
\to f_{\lambda}$ (Equation~\ref{eqn:SNRgraz}) provides an excellent approximation.
}
\label{fig:SNR_approx2}
\end{center}
\end{figure}

As shown in Fig.~\ref{fig:SNR_approx2}, this approximation provides
an excellent match to the true function with a considerably simpler
form than that of the $\lambda$ function.

As expected, Equations~(\ref{eqn:SNRin}) and (\ref{eqn:SNRgrazing})
are equivalent in the limit of $b\to(1-p)$, meaning that these two functions
join together at the boundary dividing them.

\begin{figure}
\begin{center}
\includegraphics[width=8.4 cm]{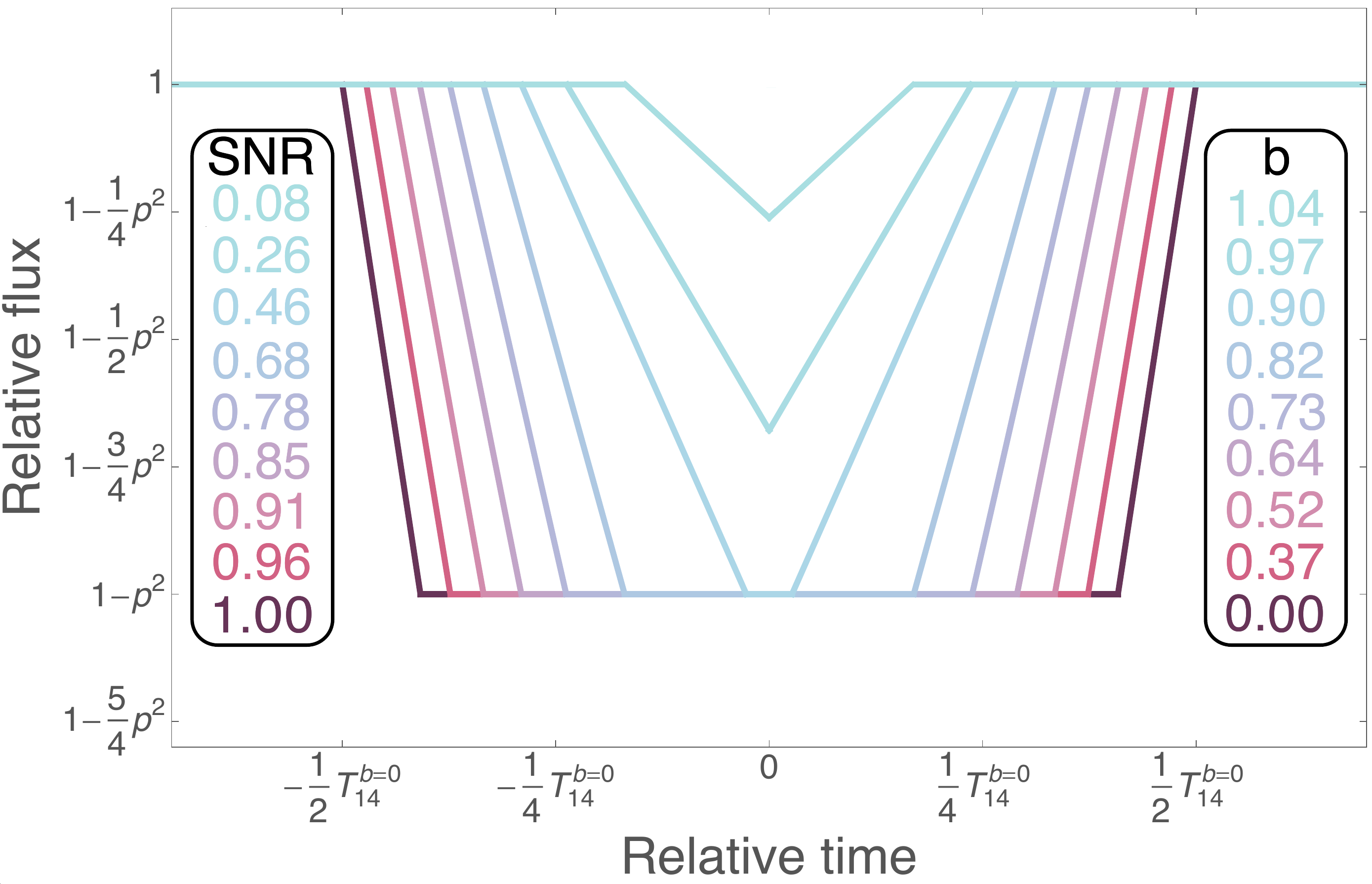}
\caption{Examples of the trapezoidal transits following our model (colored lines).
The adopted parameters are those of a typical hot Jupiter, except for
$b$, which is evenly spaced in $(1-b^2)$ space. For each example transit, we list both
the $b$ value and the associated $\SNR$ (relative to the $b=0$ case).}
\label{fig:lightcurves}
\end{center}
\end{figure}

In Fig.~\ref{fig:lightcurves}, we plot a sample of trapezoidal light curves
following our model, with the associated $b$ and $\SNR$ values listed as a
visual guide.

\subsection{Updated joint conditional probability distribution}

The joint probability distribution of the transit parameters, conditioned upon
the fact that a planet transits and the transit is detectable, is directly
proportional to the SNR, as shown in
Equations~(\ref{eqn:SNRproportional}) and (\ref{eqn:Theta_prior_fullscaling}).
As before then, the joint probability may be expressed as

\begin{align}
\pdf(\boldsymbol{\Theta}|\tdet,\tran) \propto \pdf(\tdet,\tran|\boldsymbol{\Theta}) \pdf(\boldsymbol{\Theta}).
\end{align}

The $\pdf(\tdet,\tran|\boldsymbol{\Theta})$ represents the observational bias,
which using the simpler approximation in \S\ref{sec:conditionals} was given by
Equation~(\ref{eqn:bias}). Now, the form of $\pdf(\tdet,\tran|\boldsymbol{\Theta})$
is modified to account for the trapezoidal shape of the transit, but the
only terms which have been changed relate to $p$ and $b$.

As an additional improvement, we modify the geometric bias (Equation~\ref{eqn:geometric})
to include grazing events, such that transits are now defined as $b<1+p$. Following the
methodology described in \citet{eprior:2014}, it may be
shown that this modifies $\pdf(\tran|\boldsymbol{\Theta}) \to (1+p)\pdf(\tran|\boldsymbol{\Theta})$.

Accordingly, we may express the observational bias as

\begin{equation}
\pdf(\tdet,\tran|\boldsymbol{\Theta})\propto \kappa \times \left\{ \begin{array}{ll}
\eta(p,b) & 0 \le b < 1-p \\
\zeta(p,b) & 1-p \le b < 1+p \\
0 & 1+p \le b < \infty  \end{array}
       \right. ,\label{eqn:advancedbias}
\end{equation}

where $\kappa$ is defined in Equation~(\ref{eqn:kappa}), and

\begin{align}
\eta(p,b) \equiv& \tfrac{1}{2} (1+p) ((1+p)^2 - b^2)^{-1/4} p^2 \nonumber\\
\qquad& \Big( \sqrt{(1+p)^2 - b^2} + \sqrt{(1-p)^2 - b^2} \Big),\\
\zeta(p,b) \equiv& \tfrac{1}{2} (1+p) ((1+p)^2-b^2)^{1/4} \frac{(1+p-b)^2}{4} \nonumber\\
\qquad& \Bigg( 1 - 2\log_4\Big[\frac{1+p-b}{2p}\Big] \Bigg).
\end{align}

In the above, a Taylor series expansion in small $p$ provides a first-order
leading term equal to that of Equation~(\ref{eqn:bias}), demonstrating the
equivalency of the two derivations.

\subsection{Marginalized distributions}
\label{sub:marg}

We first note that, as in the simple case, the observational bias terms are largely
separable, with the exceptions of $e$ and $\omega$, and now $p$ and $b$, being covariant:

\begin{align}
\pdf(\tdet,\tran|\boldsymbol{\Theta}) &\propto 
\pdf(\tdet,\tran|P) 
\pdf(\tdet,\tran|e,\omega)
\pdf(\tdet,\tran|b,p).
\end{align}

If we further assume that the intrinsic distributions are fully separable, then
the above simplifies the derivation of marginalized distributions, since we can
write, for example,

\begin{align}
\pdf(P|\tdet,\tran) &\propto \int \int \int \int \int \pdf(\tdet,\tran|\boldsymbol{\Theta}) \pdf(\boldsymbol{\Theta})
\,\mathrm{d}p\,\mathrm{d}b\,\mathrm{d}\rho_{\star}\,\mathrm{d}e\,\mathrm{d}\omega \nonumber\\
\qquad&\propto \pdf(\tdet,\tran|P) \pdf(P),
\end{align}

where in the case of the above, adopting a log-uniform intrinsic period
distribution ($\propto P^{-1}$), would mean that $\pdf(P|\tdet,\tran) 
\propto P^{-2}$. Repeating this marginalization for the other terms allows us to write

\begin{align}
\pdf(P|\tdet,\tran) &\propto P^{-1} \pdf(P),\\
\pdf(\rho_{\star}|\tdet,\tran) &\propto \rho_{\star}^{-1/2} \pdf(\rho_{\star}),\\
\pdf(e,\omega|\tdet,\tran) &\propto \frac{(1+e\sin\omega)^{1/2}}{(1-e^2)^{3/4}} \pdf(e) \pdf(\omega),\\
\pdf(b,p|\tdet,\tran) &\propto \pdf(\tdet,\tran|b,p) \pdf(b) \pdf(p),
\end{align}

where we do not write out the last function due to its split-domain nature (see Sections \ref{subsub:p} and \ref{subsub:b} for discussion). First, though, note that we do not impose specific intrinsic priors in the
above, allowing for flexibility in their usage. The only critical assumption in deriving the above
is that the intrinsic distributions are separable.

\subsubsection{The case of $p$}
\label{subsub:p}

In the case of the simple priors derived earlier, $p$ and $b$ are separable, in which case one can show that

\begin{align}
\pdf(b|\tdet,\tran) &\propto (1-b^2)^{1/4} \pdf(b),\\
\pdf(p|\tdet,\tran) &\propto p^2 \pdf(p).
\end{align}

The latter expression reflects our intuition that detected transits are biased towards larger
planets with a quadratic scaling. The bias evident in $\pdf(b|\tdet,\tran)$ is likely 
unfamiliar to the reader, but we show later in Fig.~\ref{fig:swift} that it provides a
good match to observations.

For the more advanced case, calculating the one-dimensional marginalized $\pdf(p|\tdet,\tran)$
and $\pdf(b|\tdet,\tran)$ can only be accomplished by assuming an intrinsic distribution for
$\pdf(p)$ and $\pdf(b)$. For impact parameter $b$, a uniform prior is reasonable and expected,
but $p$ is somewhat non-trivial. Before tackling this issue, though, the fact that we can
define $\pdf(b)$ means we can marginalize it out to derive $\pdf(p|\tdet,\tran)$. Doing so
leads to a highly elaborate expression (which cannot be compactly written out here) involving
generalized hypergeometric functions.

\begin{figure}
\begin{center}
\includegraphics[width=8.4 cm]{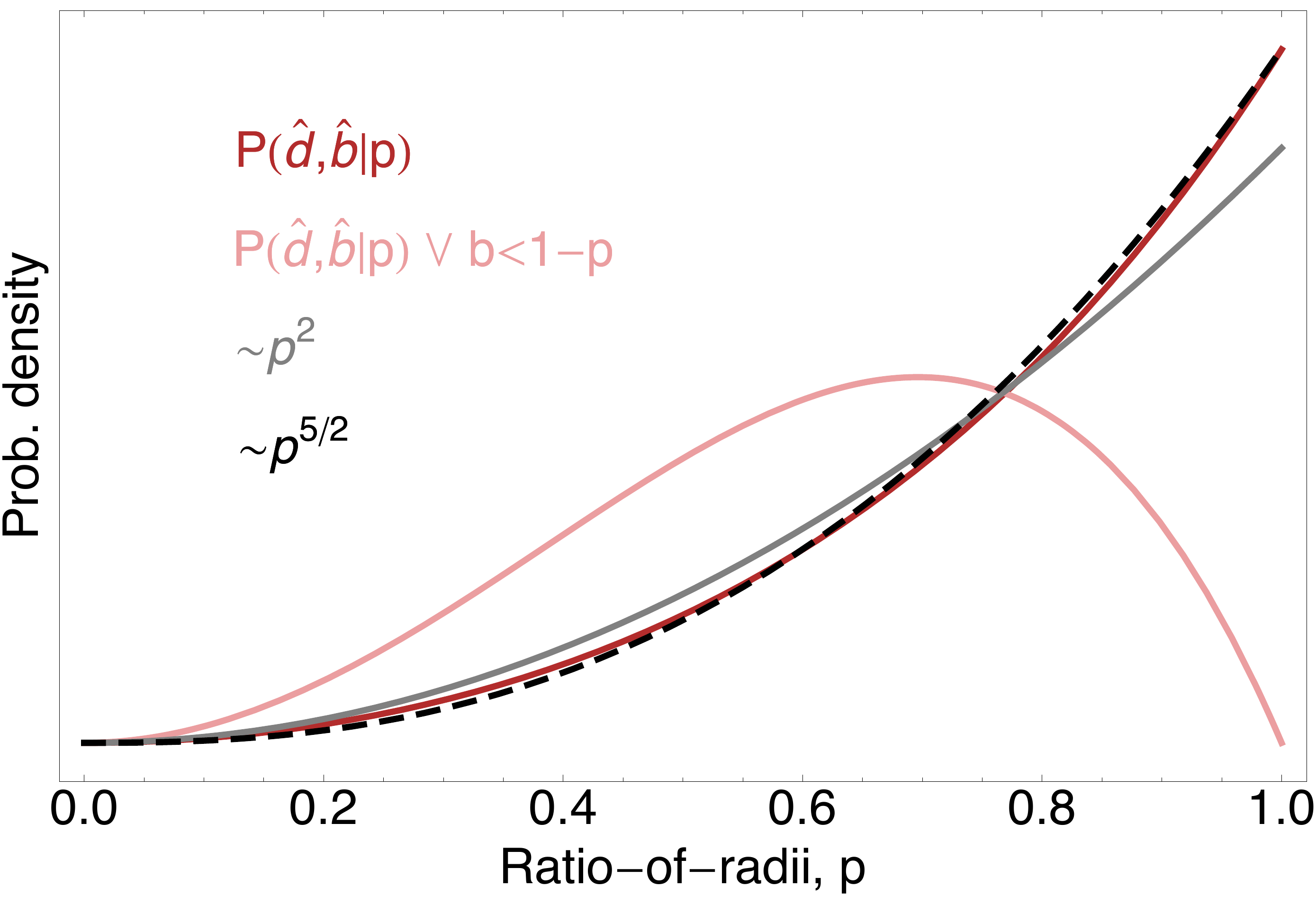}
\caption{
Red line shows the analytic (but highly elaborate) functional dependence of
$\pdf(\tdet,\tran|p)$ with respect to $p$, which is clearly super-quadratic (grey line)
and very well approximated by a power law of index $5/2$ (black line). The pink line shows the
full function in the case where V-shaped (i.e., grazing) transits are not considered `detections.'
}
\label{fig:pmarg}
\end{center}
\end{figure}

Plotting our solved function $\pdf(p|\tdet,\tran)$ (which is purely
dependent on $p$) in Fig.~\ref{fig:pmarg} reveals that the function is 
super-quadratic with respect to $p$. This implies that if one assumed a detection 
bias of $p^2$, one would underestimate the true number of small
planets, since the bias towards $p$ is actually sharper. Additionally, we observe that
the plotted function closely resembles a power-law, for which we can fit the index. We therefore
approximate

\begin{align}
\pdf(p|\tdet,\tran) &\propto p^{\alpha} \pdf(p).
\end{align}

A least-squares fit of samples suggests $\alpha=2.424$, but plotting even $\alpha=5/2$
reveals excellent agreement. It is important to stress that this value is not dependent upon
any assumption regarding the intrinsic distributions, except that impact parameter is uniformly
distributed and separable from $p$. Under this assumption, and the assumption that detectability
is proportional to SNR, geometric effects conspire to favour the detection of planets with a
scaling of $p^{5/2}$. We also tried repeating the derivation marginalizing over only $b<1-p$,
which is equivalent to insisting that V-shaped transits are not considered `detections.' This
line, plotted in pink in Fig.~\ref{fig:pmarg}, reveals a turn-over at $p=0.696$, since it is increasingly unlikely for the disc of a large planet to fully pass in front of the star.

\subsubsection{The case of $b$}
\label{subsub:b}

We are unable to derive a closed-form expression for the marginalized distribution of  
$\pdf(b|\tdet,\tran)$. Instead, we perform the integral numerically over a grid of
$b$ values in order to plot the functional form of the expression. For this calculation,
we assume that the logarithm of $p$ is distributed uniformly from $10^{-3}$ to a fixed
upper limit, which we try varying.

\begin{figure}
\begin{center}
\includegraphics[width=8.4 cm]{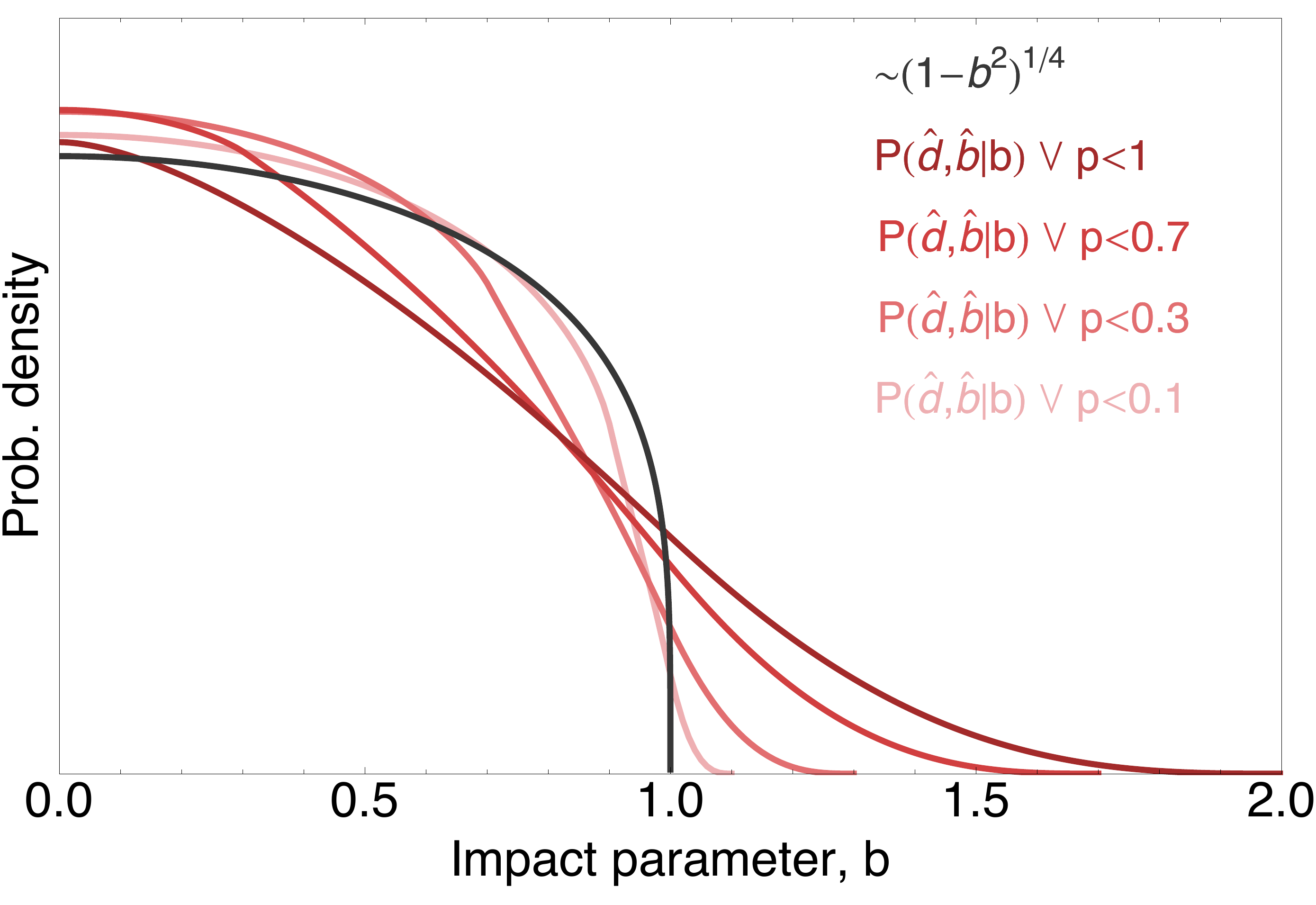}
\caption{
Coloured lines show the analytic (but highly elaborate) functional dependence of
$\pdf(\tdet,\tran|b)$ with respect to $b$, the observational bias to transit impact
parameter. The black line shows the same bias derived in the small-planet ($p\to0$) 
limit of a box-like transit.
}
\label{fig:bmarg}
\end{center}
\end{figure}

The results, shown in Fig.~\ref{fig:bmarg}, illustrate that the functional form is
generally flat at low $b$ and follows the approximate shape of the simple form
derived earlier (i.e. $(1-b^2)^{1/4}$). The effect of the $p$ marginalization largely
appears to be that of a convolution kernel, smoothing out the distribution.

\subsubsection{The case of $e$}

For eccentricity and argument of periastron, we assume that $\omega$ has a
uniform intrinsic distribution. Regardless of the intrinsic distribution for $e$, we find that
marginalizing over $\omega$ is possible analytically, as

\begin{align}
\pdf(e|\tdet,\tran) &\propto \pdf(e) \int_{\omega=0}^{2\pi} \pdf(e,\omega|\tdet,\tran) \pdf(\omega)\,\mathrm{d}\omega,
\end{align}

giving

\begin{align}
\pdf(e|\tdet,\tran) &\propto \frac{\pdf(e)}{(1-e^2)^{3/4}} \Bigg( \sqrt{1-e} E\Big[ \frac{2e}{e-1} \Big] + \sqrt{1+e} E\Big[ \frac{2e}{e+1} \Big] \Bigg),
\label{eqn:eresult}
\end{align}

where $E(x)$ is the complete elliptic integral of $x$.
The above is essentially an exponential-like function multiplied by the intrinsic $e$ prior
and thus distorts the observed eccentricity distribution towards more elliptical orbits
than the true underlying distribution.

\subsubsection{The case of $\omega$}

\begin{figure}
\begin{center}
\includegraphics[width=8.4 cm]{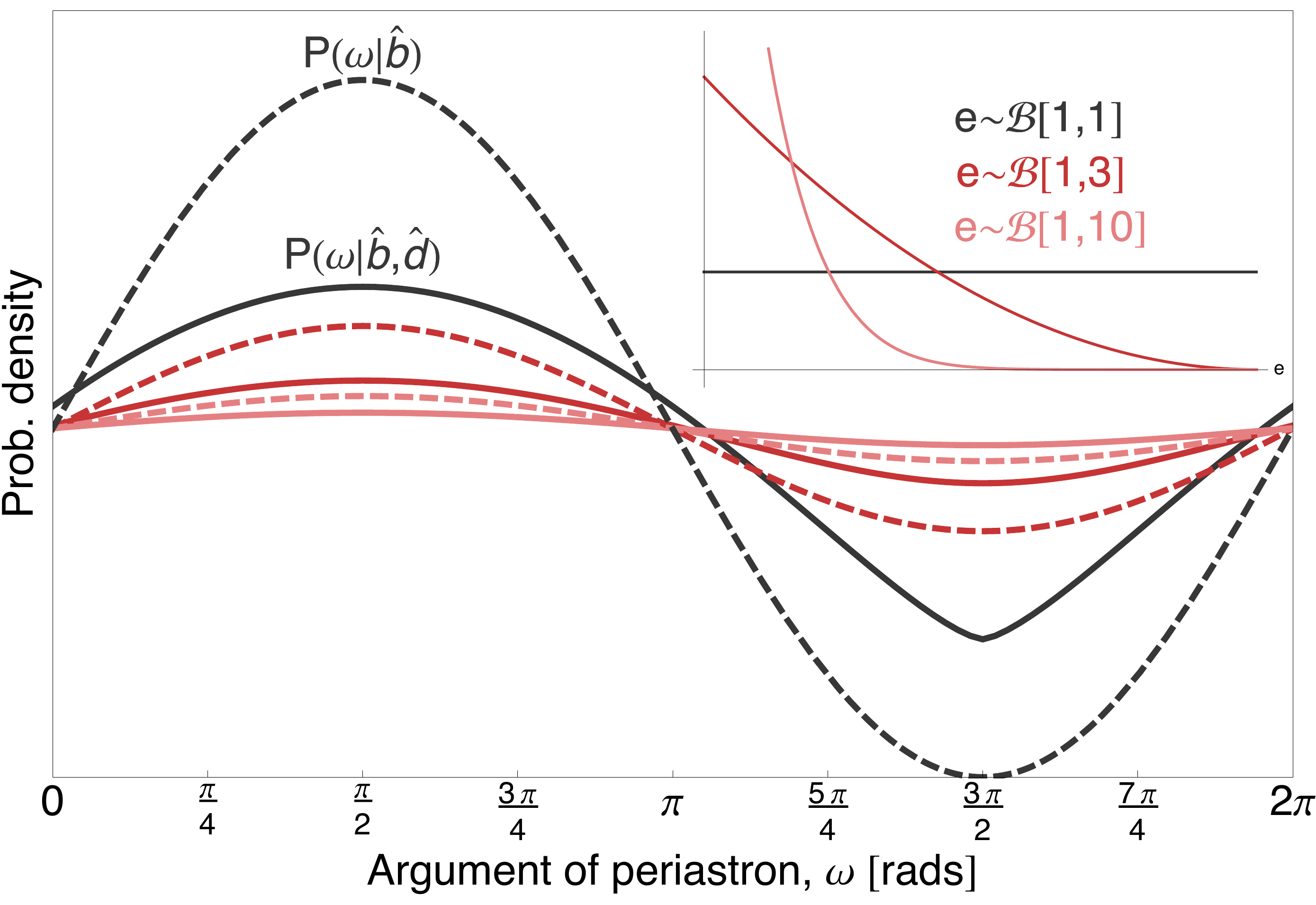}
\caption{
Solid coloured lines show $\pdf(\omega|\tdet,\tran)$, whereas dashed lines show 
$\pdf(\omega|\tran)$ for comparison, using the same inputs. Including detection
bias suppresses the strong bias towards periastron transits.
}
\label{fig:omegamarg}
\end{center}
\end{figure}

Repeating this trick for $\omega$ is more challenging and is functionally dependent upon
the assumed form for $\pdf(e)$. In what follows, we adopt a Beta distribution for $e$,
for reasons discussed in \citet{beta:2013}. Whilst a closed-form expression was not found,
numerically marginalizing over $e$ for three different choices of the Beta shape parameters
allows us to visualize the effect of the detection bias, as shown in Fig.~\ref{fig:omegamarg}.

In this figure, we compare the results of including both detection bias and geometric bias (this
work), versus the geometric bias alone (expressions of \citealt{eprior:2014}). It is clear
that our expressions significantly suppress the $\omega$ biases, leading to a population more
evenly distributed with respect to $\omega$. This can be understood by considering that apoastron
transits, whilst geometrically disfavoured, have a significant detection enhancement due to their
longer transit durations.

\section{Discussion}
\label{sec:discussion}

In this work, we have derived the joint probability distribution
of the basic transit parameters conditioned upon two observational
biases affecting this type of measurement: geometric bias and detection bias.
By treating the transit as a trapezoid, accounting for grazing events, and 
using conditional probability theory, we have derived an analytic, closed form
for these biases (see Equation~\ref{eqn:advancedbias}).

Whereas previous works have used analytic arguments to predict the detection
yields of transit surveys subject to observational biases (e.g. \citealt{beatty:2008}), the focus here is the
distortion of the observed transit parameter distributions away from those of the underlying exoplanet
population. Inferences about the properties of \Kepler\ planets are strongly
affected by geometric and detection bias, and our work provides a general framework to 
interpret the observed trends.

Our work is not intended to replace numerical Monte Carlo simulations, which
typically perform injection/recovery tests through a detection pipeline (e.g.
see \citealt{petigura:2013}; \citealt{christiansen:2015}; 
\citealt{dressing:2015}). These results, tailored to a specific mission,
have both advantages in their flexibility and disadvantages in obscuring the
mechanisms responsible for various trends. In this sense, our work
complements numerical efforts by providing insight into what observational
trends are inherent to the transit technique, rather than effects localized
to a particular survey. We highlight here several important results from our
work.

\begin{figure}
\begin{center}
\includegraphics[width=8.4 cm]{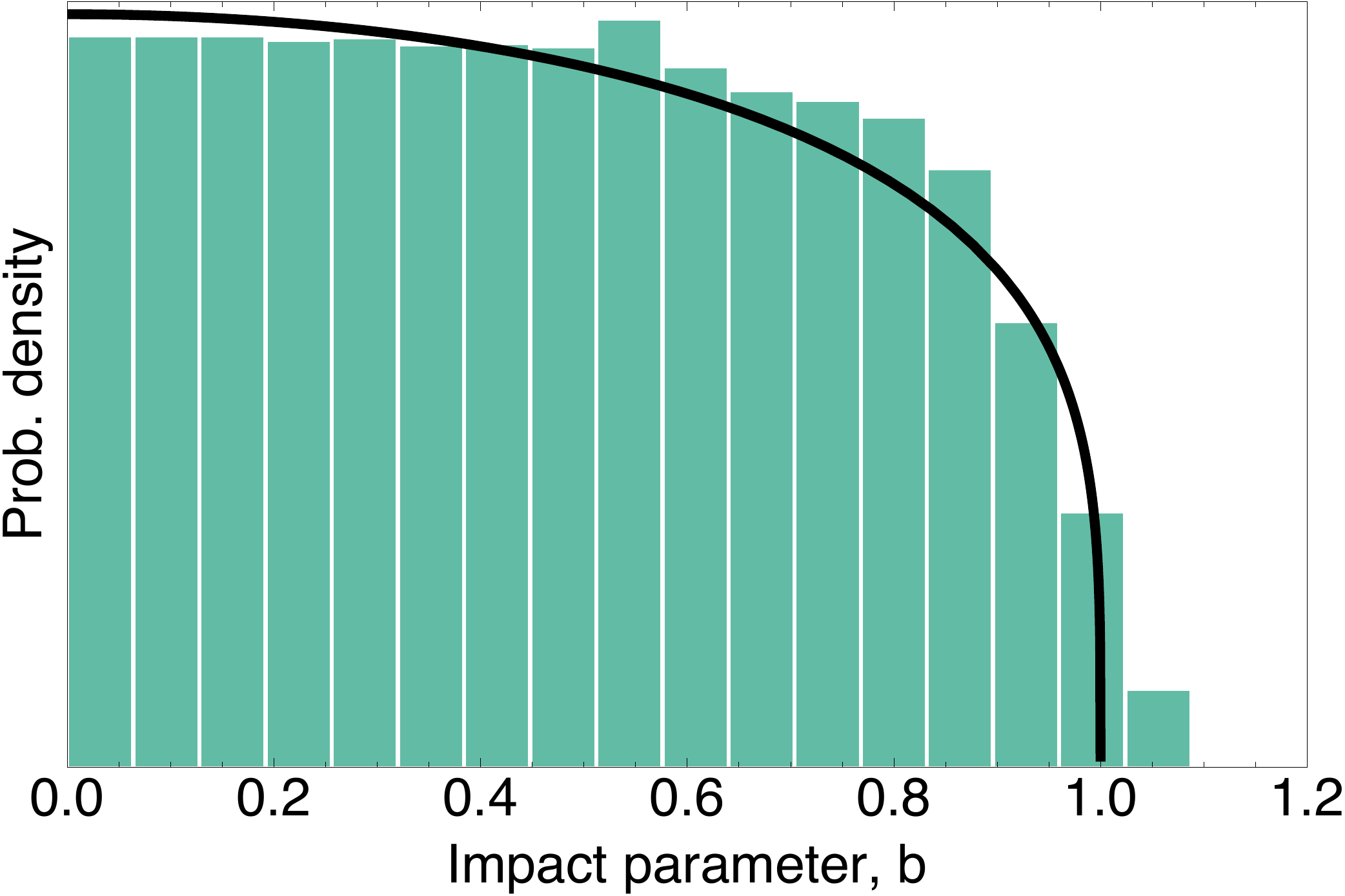}
\caption{Distribution of the transit impact parameter for the sample 
of \Kepler\ planetary candidates studied in \citet{swift:2015}. The 
measured distribution (green) is reasonably well approximated by the 
(simple case) prior $\pdf(b|\tdet,\tran)$ of Equation~(\ref{eqn:bias}) 
derived in this work.}
\label{fig:swift}
\end{center}
\end{figure}

We predict that observational bias should lead to a non-uniform distribution
in the impact parameters of detected transiting planets. This prediction may
be tested with the \Kepler\ catalogue, for which we turn to \citet{swift:2015}, who 
derived homogeneous posterior distributions for the basic transit parameters, 
including $b$, of 163 \Kepler\ planetary candidates. The cumulative 
distributions of these terms are shown in fig.~9 of that work, of which we 
reproduce a version in Fig.~\ref{fig:swift} here.

Despite the fact that \citet{swift:2015} impose a uniform prior
on $b$, the overall distribution drops off at high $b$,
reproducing our expectation that such cases are indeed
less frequent in a SNR-limited survey like
\Kepler. Fig.~\ref{fig:swift} compares the measured
distribution to that predicted by our function $\pdf(b|\tdet,\tran)$
in the simple case\footnote{Since the more advanced distribution displays
covariance between $b$ and $p$, we cannot plot it without knowledge
of the intrinsic $p$ distribution.} of Equation~(\ref{eqn:bias}), 
which displays a reasonable match to the observed shape.

Aside from providing an explanation for the distribution of impact 
parameters observed in real data \citep{swift:2015}, we highlight two other 
important results from our work. First, we update the well-known eccentricity bias of 
transits (\citealt{barnes:2007}; \citealt{burke:2008}; \citealt{eprior:2014})
to include detection bias. We find that the previously reported
bias of transiting planets towards being preferentially near periastron is substantially 
relaxed by virtue of near-apoastron transits being much longer and thus more 
detectable. Equation~(\ref{eqn:eresult}) provides a general formulation
to correct for the eccentricity bias for any assumed intrinsic distribution.

Finally, we highlight that we find the observational bias of transits with respect
to the ratio-of-radii, $p$, is not quadratic (i.e. $p^2$), as
commonly assumed, but actually super-quadratic and well-approximated by $p^{5/2}$ (see 
Fig.~\ref{fig:pmarg} and Section~\ref{subsub:p}). Critically, this result is
independent of any assumption about the intrinsic distribution of $p$ and only
assumes that impact parameter is intrinsically uniformly distributed. This result is driven 
by larger planets having (i) a longer transit duration and (ii) a higher geometric transit 
probability. Excluding grazing events (i.e. assuming $b<1-p$) leads to a non-monotonic
form for the bias with respect to $p$. These results highlight the importance of 
correctly accounting for observational biases in statistical calculations seeking
to infer the true planetary radius distribution.
	
\section*{Acknowledgements}

This research has made use of the Exoplanet Orbit Database
and the Exoplanet Data Explorer at 
\href{http://www.exoplanets.org}{exoplanets.org},
and the {\tt corner.py} code by Dan
Foreman-Mackey at 
\href{http://github.com/dfm/corner.py}{github.com/dfm/corner.py}.

\appendix

\section{An Approximation for $\lambda$}
\label{app:lambda}

The expression for $\lambda$ (Equation~\ref{eqn:lambda}) is restrictive for
further analytic work, and thus we seek an approximation via a simple polynomial,
aiming to capture both the curvature and boundary conditions of the function. 
We consider using a so-called smooth-step function, $f[x]$, over the interval
$0<x<1$, where $f[x]$ varies smoothly from $f[0]=0$ to $f[1]=1$. The function 
$\lambda$ is approximated by

\begin{align}
\lambda &\simeq p^2 f[ \mathcal{S}' ],\nonumber\\
\mathcal{S}' &= \frac{1 + p - \mathcal{S}}{2 p}.
\end{align}

The simplest function we might propose is $f_1[x] = x$, which is simply a straight
line. A common smooth-step function used in computer graphics is

\begin{align}
f_2[x] = 3 x^2 - 2 x^3.
\end{align}

Several other smooth-step functions exist at higher order, such as

\begin{align}
f_3[x] &= 6 x^5 - 15 x^4 + 10 x^3,\\
f_4[x] &= -20 x^7 + 70 x^6 - 84 x^5 + 35 x^4.
\end{align}

\begin{figure}
\begin{center}
\includegraphics[width=8.4 cm]{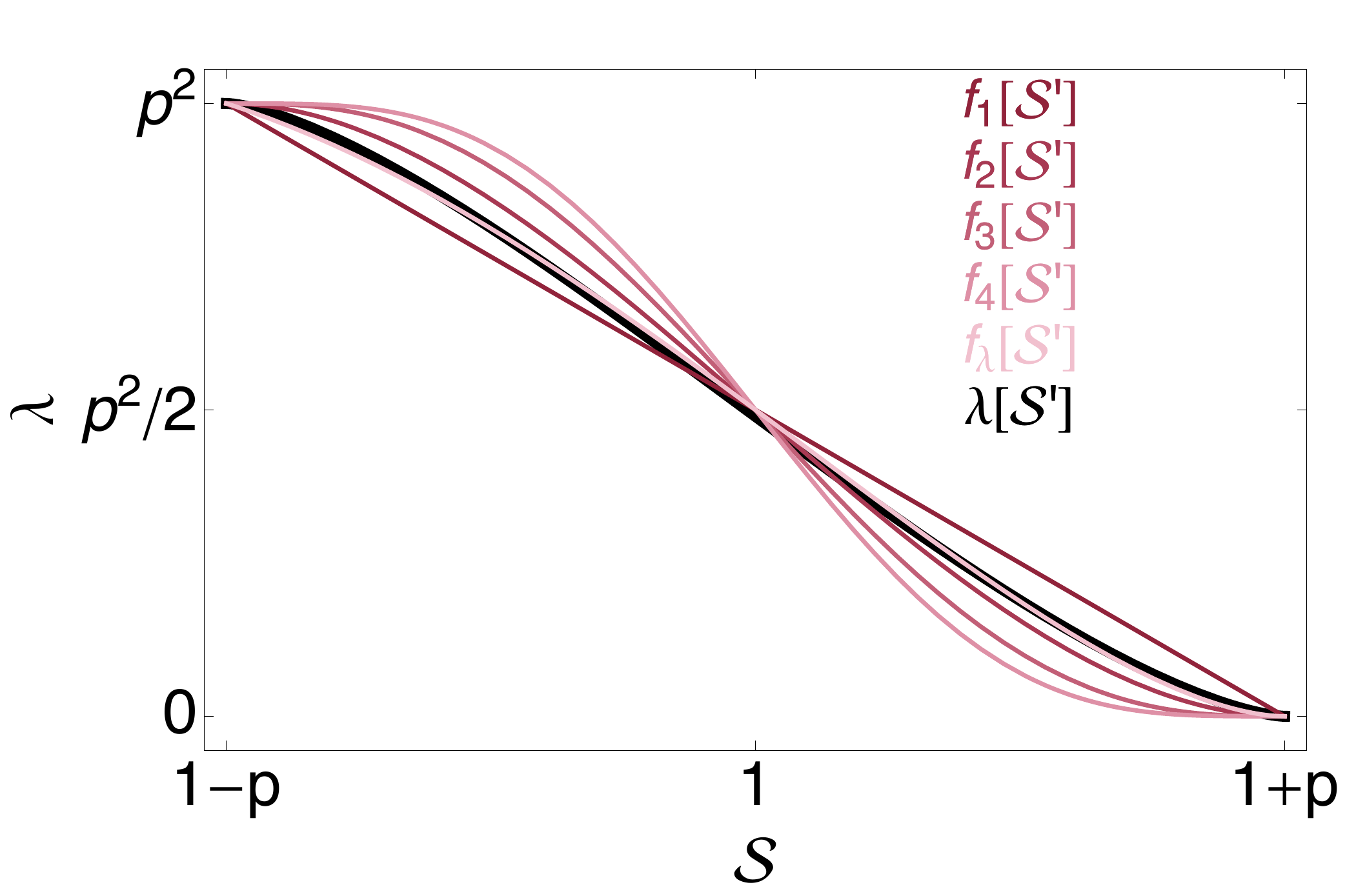}
\caption{
Comparison of several ``smooth-step'' approximations to the
$\lambda$ function. Our newly derived function, $f_{\lambda}$, provides an
excellent match.
}
\label{fig:lambda_approx}
\end{center}
\end{figure}

Plotting these functions in Fig.~\ref{fig:lambda_approx}, it apparent that
the true function, $\lambda$, lies between $f_1[\mathcal{S}']$ and
$f_2[\mathcal{S}']$. We therefore seek to generalize the $f_2[\mathcal{S}']$
expression by writing:

\begin{align}
f_{\lambda}[x] = (\alpha+1) x^{\beta} - \alpha x^{\gamma}.
\end{align}

We note that the $\lambda$ function intersects $\frac{1}{2}$ for
$\lim_{p\to0} \lambda(p,1) = \frac{1}{2}$, which implies that
we expect $f_{\lambda}[\frac{1}{2}] = \frac{1}{2}$. Imposing the above
and solving for $\alpha$ allows us to replace

\begin{align}
\alpha &= - \frac{ 2^{\gamma-1} (2^{\beta} - 2) }{ 2^{\beta} - 2^{\gamma} }.
\end{align}

Experimenting with different low-integer combinations of $\beta$ and $\gamma$, we find that setting $\beta=\gamma=2$ yields a close
match. Indeed, setting $\beta=\gamma$ and applying a least squares fit for
$\beta$ yields $\beta = 2$. Adopting this value, we therefore use the following
approximate expression for $\lambda$ (plotted in Fig.~\ref{fig:lambda_approx}):

\begin{align}
f_{\lambda}[x] &= x^2 ( 1 - \log_4 x^2 ).
\label{eqn:flambda}
\end{align}

\bsp
\label{lastpage}

\begin{thebibliography}{99}
\bibitem[\protect\citeauthoryear{Barnes}{2007}]{barnes:2007} 
Barnes, J.~W., 2007, PASP, 119, 986
\bibitem[\protect\citeauthoryear{Beatty \& Gaudi}{2008}]{beatty:2008} 
Beatty, T.~G. \& Gaudi, S.~B., 2008, ApJ, 686, 1302
\bibitem[\protect\citeauthoryear{Burke}{2008}]{burke:2008} 
Burke, C.~J., 2008, ApJ, 679, 1566
\bibitem[\protect\citeauthoryear{Burke et al.}{2015}]{burke:2015} 
Burke, C.~J. et al., 2015, ApJ, 809, 8
\bibitem[\protect\citeauthoryear{Carter et al.}{2008}]{carter:2008} 
Carter, J.~A., Yee, J.~C., Eastman, J., Scott, S.~B. \& Winn, J.~N., 2008, ApJ, 689, 499
\bibitem[\protect\citeauthoryear{Christiansen et al.}{2015}]{christiansen:2015} 
Christiansen, J.~L. et al., 2015, ApJ, 810, 95
\bibitem[\protect\citeauthoryear{Dalcanton et al.}{2015}]{dalcanton:2015} 
Dalcanton, J. et al., 2015, preprint (arXiv:1507.04779)
\bibitem[\protect\citeauthoryear{Dressing \& Charbonneau}{2015}]{dressing:2015} 
Dressing, C.~D. \& Charbonneau, D., 2015, ApJ, 807, 45
\bibitem[\protect\citeauthoryear{Fressin et al.}{2013}]{fressin:2013} 
Fressin, F. et al., 2013, ApJ, 766, 81
\bibitem[\protect\citeauthoryear{Jenkins et al.}{2010}]{jenkins:2010} 
Jenkins, J.~M. et al., 2010, ApJ, 713, 87
\bibitem[\protect\citeauthoryear{Kipping}{2010}]{investigations:2010} 
Kipping, D.~M., 2010, MNRAS, 407, 301
\bibitem[\protect\citeauthoryear{Kipping}{2013}]{beta:2013} 
Kipping, D.~M., 2013, MNRAS, 434, L51
\bibitem[\protect\citeauthoryear{Kipping}{2014}]{eprior:2014} 
Kipping, D.~M., 2014, MNRAS, 444, 2263
\bibitem[\protect\citeauthoryear{Mandel \& Agol}{2002}]{mandel:2002} 
Mandel, K. \& Agol, E., 2002, ApJ, 580, 171
\bibitem[\protect\citeauthoryear{Petigura et al.}{2013}]{petigura:2013} 
Petigura, E.~A., Howard, A.~W. \& Marcy, G.~W., 2013, PNAS, 110, 19273
\bibitem[\protect\citeauthoryear{Swift et al.}{2015}]{swift:2015} 
Swift, J.~J., Montet, B.~T., Vanderburg, A., Morton, T., Muirhead, P.~S. \&
Johnson, J.~A., 2015, ApJ, 218, 26
\end{thebibliography}
\end{document}